\pdfoutput=1
\documentclass[runningheads]{llncs}
\usepackage{graphicx}
\usepackage{amsmath,amssymb} % define this before the line numbering.
\usepackage{float}
\usepackage{color}

\usepackage[T1]{fontenc}
\usepackage[utf8]{inputenc}

\begin{document}
\title{UNDERSTANDING AND CLASSIFYING CULTURAL MUSIC USING MELODIC FEATURES \newline
Case of Hindustani, Carnatic and Turkish Music } 
% STYLE CLASSIFICATION IN EASTERN CULTURAL MUSIC USING DISCRIMINATORY MELODIC FEATURES

\titlerunning{STYLE CLASSIFICATION  USING MELODIC FEATURES}
% Replace with a meaningful short version of your title
%
\author{Amruta Vidwans,
Prateek Verma, 
Preeti Rao
}

%
%Please write out author names in full in the paper, i.e. full given and family names. 
%If any authors have names that can be parsed into FirstName LastName in multiple ways, please include the correct parsing, in a comment to the volume editors:
%\index{Lastnames, Firstnames}
%(Do not uncomment it, because you may introduce extra index items if you do that, we will use scripts for introducing index entries...)
\authorrunning{Amruta Vidwans, Prateek Verma and Preeti Rao}
% Replace with shorter version of the author list. If there are more authors than fits a line, please use A. Author et al.
%

\institute{Digital Audio Processing Lab, Department of Electrical Engineering, \linebreak
Indian Institute of Technology, Bombay, 400076, India.\linebreak
\email{\{amrutav,prateekv,prao\}@ee.iitb.ac.in}}
\maketitle              % typeset the header of the contribution

\begin{abstract}
\footnote{The work appeared in the 3rd CompMusic Workshop  "Developing Computational models for the Discovery of the World’s Music" held at IIT Madras, Chennai, 2013}
%\texttt{\url{https://compmusic.upf.edu/node/189}}}

We present melody based classification of musical styles by exploiting the pitch and energy based characteristics derived from the audio signal. Three prominent musical styles were chosen which have improvisation as integral part with similar melodic principles, theme, and structure of concerts namely, Hindustani, Carnatic and Turkish music. Listeners of one or more of these genres can discriminate between these based on the melodic contour alone. Listening tests were carried out using melodic attributes alone, on similar melodic pieces with respect to \textit{raga}/\textit{makam}, and removing any instrumentation cue to validate our hypothesis that style distinction is evident in the melody. Our method is based on finding a set of highly discriminatory features, derived from musicology, to capture distinct characteristics of the melodic contour. Behavior in terms of transitions of the pitch contour, the presence of micro-tonal notes and the nature of variations in the vocal energy are exploited. The automatically classified style labels are found to correlate well with subjective listening judgments. This was verified by using statistical tests to compare the labels from subjective and objective judgments. The melody based features, when combined with timbre based features, were seen to improve the classification performance.

\end{abstract}

\section{Introduction}
Indian classical music is mainly categorized into two styles, viz. Hindustani and Carnatic. They have similar \textit{raga} (melodic mode) and \textit{tala}(rhythm) framework while they differ in the performance structure, melodic movements and also in the type of instruments used. Similar to Indian classical music, in terms of the \textit{raga} framework, is Turkish music which has \textit{makam} i.e. a scale and melodic movements associated with it. The structure of the concert is similar as well in the three music styles with the concert starting with an unmetered section called \textit{alap}/\textit{taqsim} section and further proceeding into the improvisations in the metered section adhering to the \textit{raga}/\textit{makam} rules and a concept of tonic.
There has been some past work on Indian classical music on motif identification~\cite{ross2012detecting}, \textit{raga} recognition \cite{chordia2007automatic}\cite{koduri2011survey} and sub-genre classification~\cite{kini2011automatic}. Classification for different styles viz. Arabic, Chinese, Japanese, Indian, African, Western Classical music using timbre, rhythm and wavelet coefficients of decomposing the audio signal via multi-level Daubechies wavelet was proposed in ~\cite{liu2009cultural}. They emphasized that the diversity existing within Indian classical music is difficult to model. Audio pieces were divided into 9 world regions using rhythmic, tonal and timbral features with new features for non-western music in~\cite{kruspe2011automatic}. Melody based classification for western music styles was done by obtaining high-level melodic descriptors that could be easily related to the properties of the music~\cite{salamon2012musical}. Features based on timbre such as MFCC, delta-MFCC, spectral features etc. borrowed from speech processing were used in~\cite{agarwal2013comparative} to distinguish different Indian music genres rather than using any specific musical attributes. 
\par
The present work builds upon our previous work~\cite{vidwans2012classification} that explored features, motivated from musicology, based on local characteristics of the melodic contour for the automatic classification of Indian classical vocal styles into Hindustani and Carnatic. New features that improve upon the discrimination by including features at larger time-scales than previous are proposed here and evaluated on a larger dataset. Subjective classification results are presented together with the prediction of listener judgments by the automatic classifier.
Our objective is to address the problem of style classification of these culture specific music styles by considering melody alone. Even though there exist obvious timbre features based on instruments used (such as violin in the Carnatic style and harmonium in Hindustani concerts), and language cues in vocal concerts, we choose to discriminate the three styles using pitch attributes alone. This was done after validating via listening tests that the listeners can successfully distinguish the three styles using only a few pitch related attributes. The existing work on style classification has been improved, keeping in mind the applicability to a large dataset. The present technique can be extended in general to find the cues with which listeners distinguish musical pieces by correlating their responses with the classifier output. To the best of our knowledge, no such study of using pitch attributes alone and showing a strong correlation with the listening tests of those attributes has been done in the past. Also, the study will be applicable in recommendation systems, automatic metadata extraction, and tagging or classification databases.

\section{Database}
Concert recording CDs of widely performed \textit{ragas}/\textit{makams} by well-known artists (of various schools of music) of Hindustani, Carnatic and Turkish style were obtained and the audio was converted to 16 kHz, mono sampled at 16 bits/sample. A \textit{raga}/\textit{makam} can be defined as a scale of notes having a particular arrangement and melodic movements~\cite{itcWebsite}\cite{WikiMakam}. In these styles of music, there is always an attempt to follow a melodic structure which adheres to the rules of \textit{raga}/\textit{makam}. In the present study, we consider \textit{raga}s/\textit{makam}s that use almost the same scale intervals (relative to the chosen tonic note) in the Hindustani, Carnatic and Turkish styles. There are a total of 180 distinct concert \textit{alap}s/\textit{taqsim} (30-40s each) equally distributed across styles. The \textit{alap}/\textit{taqsim} section is a melodic improvisation that is rendered generally always at the start of each concert~\cite{WikiAlap}~\cite{WikiTaqsim} and hence it is expected to be enough to make a decision for a track by calculating features for this section. Same scale \textit{raga}s/\textit{makam}s are chosen to avoid the bias of listeners to a particular style for listening tests conducted as described in the next section. The artist information and \textit{raga}/\textit{makam} information is provided in Table \ref{tab:ArtisteNames} and \ref{tab:RagaDist} respectively. The instrument accompaniment in the \textit{alap}/\textit{taqsim} section is usually \textit{tanpura} and harmonium for Hindustani, \textit{tanpura} and violin for Carnatic and \textit{tambur}, \textit{kanun}, \textit{davul} for Turkish music.

% \usepackage{graphicx}
\begin{table}[]
\centering
\resizebox{\textwidth}{!}{%
\begin{tabular}{|l|l|l|}
\hline
\textbf{Hindustani Artistes}        & \textbf{Carnatic Artistes}                        & \textbf {Turkish Artistes}                     \\ \hline
Ajoy Chakrabarty, Aslam Khan        & A R Iyengar, G N Balasubramaniam                  & Hafız Ahmed Bey, Aaron Kohen         \\ \hline
Girija Devi, Jasraj, Kaivalya Kumar & M Balamuralikrishna, M D Ramanathan               & Mustafa Zeki Çağlarman, Hafız Osman  \\ \hline
Kumar Gandharva, Malini Rajurkar    & M S Subhalakshmi, Sangeetha Shivakumar            & Hafız Hüseyin Hüsnü, Sâdettin Kaynak \\ \hline
Shubha Mudgal, Ulhas Kashalkar      & Sanjay Subramanium, N Santanagopalan              & Hafız Kemal Bey, Yaşar Okur          \\ \hline
Fateh Ali Khan, Kishori Amonkar     & Shankaranarayanan, Sudha Raghunathan              & Hafız Memduh, Tanburi Cemil Bey      \\ \hline
Veena Sahasrabuddhe, Bhimsen Joshi  & T R Subramaniam, T S Kalyanaraman                 & Victoria Hazan                       \\ \hline
Rashid Khan, Prabha Atre            & R Vedavalli, Ramnad Krishnan, T N Seshagopalan    &                                      \\ \hline
                                    & K V Narayanswamy, Semmangudi S Iyer               &                                      \\ \hline
                                    & M L Vasanthakumari, T S Sathyavati, T. M. Krishna &                                      \\ \hline
\end{tabular}%
}
\caption{List of artists covered in the \textit{alap}/\textit{taqsim} database for Hindustani, Carnatic and Turkish styles of music}
\label{tab:ArtisteNames}
\end{table}

% \usepackage{graphicx}
\begin{table*}[]
\centering
\begin{tabular}{|c|c|c|}
\hline
\textbf{\begin{tabular}[c]{@{}l@{}}Hindustani \textit{raga}\\ (No. of clips)\end{tabular}} & \textbf{\begin{tabular}[c]{@{}l@{}}Carnatic Raga\\ (No. of clips )\end{tabular}} & \textbf{\begin{tabular}[c]{@{}l@{}}Turkish Makam\\ (No. of clips)\end{tabular}} \\ \hline
Todi (12)                                                                         & Subhapanthuvarali (14)                                                           & Nihavent (3)                                                                    \\ \hline
Malkauns (18)                                                                     & Hindolam (12)                                                                    & Rast (19), Mahur(13)                                                            \\ \hline
Jaijaiwanti (10)                                                                  & Dwijavanthy (14)                                                                 & Hüzzam (8), Neva (4)                                                            \\ \hline
Yaman and Yaman Kalyan (20)                                                       & Kalyani (20)                                                                     & Saba (10), Hüseyni (3)                                                          \\ \hline
\end{tabular}
\caption{Distribution of \textit{alap}/\textit{taqsim} clips across \textit{raga}s for automatic classification}
\label{tab:RagaDist}
\end{table*}

\section{Perception Testing}
To validate our hypothesis that melody is sufficient to capture the style distinction, we carried out listening tests. To avoid any bias towards artist identity, voice quality, and pronunciation, the melodic contour is re-synthesized using a uniform timbre vowel-like sound using 3 harmonics of equal strength before being presented to listeners. The pitch tracking was carried out using~\cite{rao2010vocal}, and manually corrected by listening to the original and re-synthesized tracks to avoid any pitch error based bias. The amplitude of the re-synthesized tone, however, follows that of the singer’s voice. Also to remove bias towards a particular \textit{raga}/\textit{makam},  \textit{raga} clips with same structure and notes in all the three styles were chosen. The amplitude is obtained by summing the energies of the vocal harmonics estimated from the detected pitch. The volume dynamics are retained together with pitch dynamics since they play a role in melody perception.

% \usepackage{graphicx}
\begin{table*}[]
\centering
\begin{tabular}{|c|c|c|c|}
\hline
\textbf{No.} & \textbf{Category}           & \textbf{No. of tests per category} & \textbf{Accuracy (\%)} \\ \hline
1            & Trained \textless{}3yrs     & 10                                    & 66                     \\ \hline
2            & Trained 3-10yrs             & 21                                    & 79                     \\ \hline
3            & Trained \textgreater{}10yrs & 7                                     & 74                     \\ \hline
4            & Avid Listener               & 39                                    & 67                     \\ \hline
5            & Amateur                     & 63                                    & 60                     \\ \hline
\end{tabular}
\caption{Listening test results for different training categories of participants}
\label{tab:ListeningTest}
\end{table*}

A total of 140 listeners were asked to participate in subjective listening experiment conducted via an online interface where 127 listeners participated with the assurance that the top 2 people would get a prize. The criterion was to achieve the best possible accuracy in distinguishing the three categories. Giving such motivation resulted in the listeners trying hard to give the best performance and avoid sub-par performances of crowd-based testing which can occur in interfaces such as Amazon Mechanical Turk. Training levels of the listeners can be seen in Table \ref{tab:ListeningTest}. 
\par
\begin{figure}
\includegraphics[width=\textwidth]{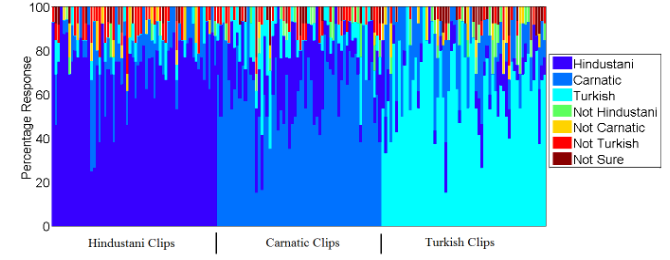}
\caption{Percentage distribution of labels for each clip marked across all the participants showing the 7 labels marked.} \label{fig1}
\end{figure}

The dataset of 180 \textit{alap} clips was divided into 10 sets. Each set contained 6 clips of each style from the dataset with almost equal number of clips for each \textit{raga}. The listeners had to listen to at least first 10s of a clip before marking their decision as Hindustani (H), Carnatic (C), Turkish (T), Not Hindustani (NH), Not Carnatic (NC), Not Turkish (NT) or Not Sure (NS). Option for a clip to play again or pause was available but skipping a clip was not allowed. We observe that listeners are able to identify the style with an average accuracy of 69\% (Correctly identifying the clip as H, C or T) with category wise accuracy as seen in Table \ref{tab:ListeningTest} and the confusion matrix as seen in Table \ref{tab:ListeningTestCM}. The decisions made across all the listeners for each clip can be seen in Figure \ref{fig1}.

% \usepackage{multirow}
% \usepackage{graphicx}
\begin{table*}[]

\centering
\begin{tabular}{|c|c|c|c|c|c|c|c|}
\hline
%\multirow
{\textbf{\begin{tabular}[c]{@{}c@{}}Ground\\ Truth\end{tabular}}} & \multicolumn{7}{c|}{\textbf{Listening Test}}                                                       \\ \cline{2-8} 
                                                                                 & \textbf{H}   & \textbf{C}   & \textbf{T}   & \textbf{NH} & \textbf{NC} & \textbf{NT} & \textbf{NS} \\ \hline
\textbf{H}                                                                       & \textbf{612} & 107          & 29           & 7           & 22          & 50          & 13          \\ \hline
\textbf{C}                                                                       & 166          & \textbf{521} & 73           & 21          & 3           & 37          & 19          \\ \hline
\textbf{T}                                                                       & 61           & 143          & \textbf{538} & 31          & 22          & 17          & 28          \\ \hline
\end{tabular}
\caption{Confusion matrix of 840 labels obtained across the listening test labels of each style (with an average of 14 decisions per clip) showing clearly the high values (boldfaced) for correct classification of each category. Note the H-C and C-T confusion is seen more as compared to H-T.}
\label{tab:ListeningTestCM}
\end{table*}

\begin{figure}
\includegraphics[width=\textwidth]{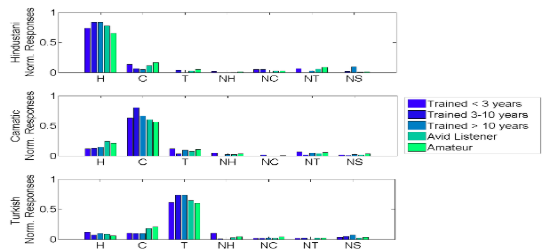}
\caption{Normalized category wise response for 60 clips of each style. The dependence on the training can be seen on the markings. } \label{fig2}
\end{figure}

Mainly the trained category participants (trained in either of Hindustani or Carnatic Music) were familiar with H or C but none of them were trained in T. It was also observed that more the music training of the listener in any style, higher is their performance to discriminate between the styles. The accuracy is comparable for trained >10yrs and trained 3-10yrs category but may have been higher for trained >10yrs category if more number of test subjects were present in it. The clips where the trained listeners with experience >10years failed were truly confusing as they had mixed characteristics of both the styles for e.g. Hindustani clip having a lot of \textit{gamak} (rapid oscillatory pitch movement) which is typically seen in the Carnatic style. Participants were able to classify H better than C and T. The confusions between H-C and C-T are comparable, with the confusions between H-T being least. The amateurs and avid listeners had more confusion in H-C than in C-T which was seen from NT option being marked more as compared to NC and NH. Also, the wrong eliminations are very few or negligible in all the listener categories, even in case of the amateur category where the participants were not familiar to any of the music styles, and this can be seen from Figure \ref{fig2}.

\section{Melodic Feature Extraction}
Melody is defined by Poliner et al.~\cite{poliner2007melody} as "the single (monophonic) pitch sequence that a listener might reproduce if asked to whistle or hum a piece of polyphonic music, and that a listener would recognize as being the ‘essence’ of that music when heard in comparison.” It is the property which differentiates sounds of the same timbre and loudness. The melody can be extracted from the audio by using melody extraction algorithms ~\cite{poliner2007melody}\cite{pant2010melody}\cite{salamon2012melody}, and represented as a time series of pitch values. This, when re-synthesized using sinusoidal basis function weighted by the timbral envelope, will sound like the original melody. The notes present in an audio excerpt can be found by plotting a histogram of the pitch values in the melody. In addition to the pitch, the second dimension of the extracted melody is the energy corresponding to it. In case of polyphonic audio, energy value corresponding to only vocal melody can be obtained by summation of energy values at harmonics of the predominant pitch.
\subsection{Preprocessing: Melody extraction}
The database in our case consists of \textit{alap} and \textit{taqsim} recordings which do not contain any percussive accompaniment. It consists of background drone and melodic accompaniment. The state of the art automatic pitch trackers have achieved an accuracy of 80\%~\cite{rao2010vocal}. The limiting factors for these algorithms in Indian and Turkish music is tracking melody with rapid pitch fluctuations, melodic accompaniment and voicing detection~\cite{rao2012signal}\cite{rao2011context}. This prompted us to use a semi-automatic approach for pitch detection~\cite{pant2010melody}. The detected pitches obtained at 10ms intervals are converted to the relative cents scale. This study will give us a fair indication of the best performance that can be achieved by our set of features.
\begin{figure}
\includegraphics[width=\textwidth]{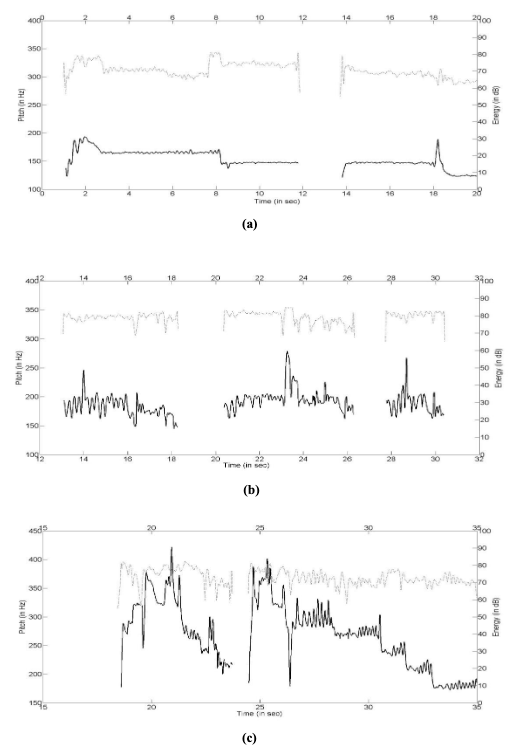}
\caption{Pitch contours of Hindustani, Carnatic and Turkish style music. The proportion of steady notes is seen more in Hindustani followed by Turkish and least in Carnatic. The dotted contours show the corresponding energy variations.
} \label{fig3}
\end{figure}

\subsection{Localized contour-shape features}
The melodic contour in Indian classical vocal styles comprise steady and ornamented (\textit{gamak}) regions. These localized characteristics were used to design discriminative features to classify Hindustani and Carnatic styles~\cite{vidwans2012classification}. The proportion of steady notes to the \textit{gamak}/ornaments is higher in Hindustani \textit{alap} section as opposed to that in Carnatic \textit{alap} or Turkish \textit{taqsim} as seen in Figure \ref{fig3}. This is captured by the stable note measure which is the ratio of the duration of steady notes to the total duration as implemented in~\cite{vidwans2012classification}. The regions not corresponding to the steady notes were characterized using the \textit{gamak} measure. The \textit{gamak} measure characterizes the oscillatory behavior of the pitch contour modulations.  The choice of parameters, one data-driven and other musicologically derived, were experimented with by~\cite{vidwans2012classification} to reliably identify the steady note and \textit{gamak} sections. While musicians labeled steady regions matching the parameters (N=700ms, J=10cents), the best classification accuracies led to the “data-driven” parameters (N=400ms, J=20cents). Here the N corresponds to the minimum duration perceived to be stable and J corresponds to the standard deviation of its pitch contour. 

\par
Here we consider the data-driven parameters to compute the two features for the three styles. The pitch contour segments that are labelled as non-steady (\textit{gamak}) region are analyzed for rate of pitch modulation. The Fourier spectrum of the temporal pitch trajectory, sampled every 10ms, shows clear peaks whenever the region is characterized by uniform oscillations. The energy ratio $ER$ is calculated for pitch values in 1s window with 0.5s hop by taking its Fourier transform. For calculating the $ER$ of a segment, we take the ratio of the energy of oscillations in the regions in 3-7.5 Hz, normalized by the energy in the 1-20Hz frequency region as shown below.

\begin{equation}
    ER=\frac{\sum ^{k_{7.5 Hz}}_{k_{3Hz}}|Z(k)|^2}{\sum ^{k_{20 Hz}}_{k_{1 Hz}}|Z(k)|^2}
\end{equation}

where $Z(k)$ is the DFT of the mean subtracted pitch trajectory $z(n)$, with samples at 10ms intervals, and $k_{f Hz}$ is the frequency bin corresponding to $ k$ Hz.

The percentage of $ER$ computed that cross a certain threshold serves as an indicator of the vocal style. We define  the \textit{Gamak Measure} as 
\begin{equation}
    \text{Gamak Measure} =\frac{\text{Number of ER > x}}{\text{ Total number of ER computed}}
\end{equation}

The threshold $x$ was varied from 0.1 to 0.9 to empirically set its value as 0.3 to achieve best separation between the oscillatory segments and relatively slowly varying segments. The energy ratio is seen to be high for Carnatic style ornaments while it is least for Hindustani style due to the presence of glides. The energy ratio is seen to be intermediate for Turkish style due to the presence of oscillations relatively slower than in the Carnatic style but faster than in Hindustani style as seen in Figure \ref{fig3}.

\subsection{Distance of maximum peak from tonic in an unfolded histogram}
\begin{figure}
\includegraphics[width=\textwidth]{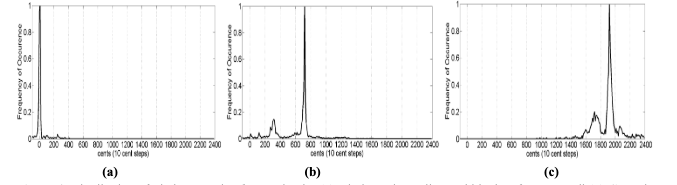}
\caption{Distribution of Pitch Range (in cents) in \textit{alap} section by (a) Hindustani vocalist Rashid Khan for \textit{raga} Todi (b) Carnatic vocalist Sudha Raghunathan for \textit{raga} Subhapanthuvarali (c)Turkish vocalist Sadettin Kaynak for \textit{makam} Mahur. \textit{Alap} is centered around 'S' in Hindustani and 'P' in Carnatic style and even higher for \textit{taqsim} section in Turkish style} \label{fig4}
\end{figure}

A study on pitch interval concentration in style perception was done by Vidwans et.al~\cite{vidwans2012classification}.  Their study pertained to finding the distance of the most occurring note rendered from the tonic for style discrimination. Their observation was that in an unfolded histogram, the Hindustani \textit{alap}s are concentrated in the region near the tonic while the Carnatic \textit{alap} pitch distribution is closer to the upper octave tonic. Further, we observed that in the case of  Turkish music that the range of the artist is predominantly in the higher octave for the \textit{taqsim} section. Hence the feature, distance of the maximum peak from tonic in the unfolded histogram, is expected to give good distinguishing feature values for the three styles. This can be observed from the feature representation of the examples chosen as seen in Figure \ref{fig4}. To find the tonic of the clips algorithm by~\cite{salamon2012multipitch} was used. 

\subsection{Melodic transitions}
\begin{figure}
\begin{center}
    \includegraphics[width=250pt]{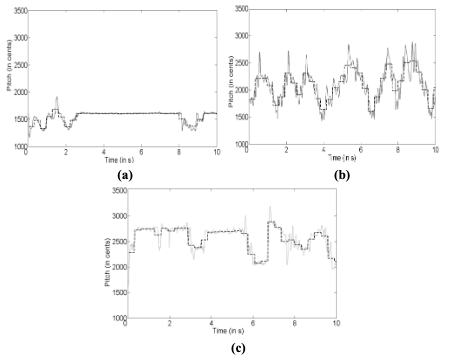}
\end{center}

\caption{Concatenated pitch contour (in gray) of (a) Carnatic and (b) Hindustani and (c) Turkish clip. The black dashed line is the 5th level Haar wavelet approximation.} \label{fig5}
\end{figure}

The feature described in section 4.2 characterizes the melody on a small time scale interval for the nature of pitch movements. The distance of highest peak from the tonic discounts the temporal nature of the pitch contour by taking the pitch histogram. For the characterization of melody, one important aspect will be to obtain a coarse representation of the melody which ignores the smaller intricate variations but yet retains the structure of the pitch contour. To obtain the desired coarse representation, one way is to use wavelet representation of the pitch contour. Wavelets have been used to model time series data~\cite{vlachos2003wavelet} capturing variations at different scales of resolution. The overall progression of the melody can be characterized by using a wavelet basis function. The Haar wavelet is suitable due to its piecewise constant nature. The hypothesis is that in case of Hindustani \textit{alap} section the singer renders 3-4 notes around the tonic in order to emphasize it, whereas in Carnatic transitions occur not necessarily around the tonic as seen in Figure \ref{fig5}. In the case of Turkish Music, the transitions are moderate as compared to Carnatic music. In order to quantify this, the vocal sections are concatenated and represented by level 5 approximation using Haar wavelet as seen in Figure \ref{fig5}. Concatenation along with Haar wavelet helps to capture the next note of the singer after the silence.

The lower levels of approximation will capture the minute details and not the overall trend and may be prone to pitch errors while the higher levels will lead to loss of information. Level 5 wavelet approximation was chosen as optimum thus giving 320ms as the minimum note duration for representation of the notes in the melody. In case of fast varying pitch movements, this may not be good but we are interested in the transitions rather than the actual note representation. The multi-level wavelet decomposition will thus give a coarse representation of the overall melody. To capture the trend, the number of upward transitions greater than 1 semitone (minimum possible jump in a \textit{raga}) in the approximated pitch contour, normalized by the length of the audio piece is taken as a feature.

\subsection{Energy based feature}
In the listening tests described in section 3, the participants also used energy as a cue to arrive at a label for a clip as given in the feedback by some of the listeners. There is a usage of tremolo in case of Turkish music while Indian music does not have tremolo which was confirmed from the clips in the database. To get an estimate of the periodicity of the energy contour, the energy contour is subtracted from its median filtered version and the number of zero crossings are used as a feature. The length of the median filter is decided empirically by considering the maximum accuracy achieved in the classification. The median filter length was kept to be 1s over energy contour values. 
\begin{figure}
\includegraphics[width=\linewidth]{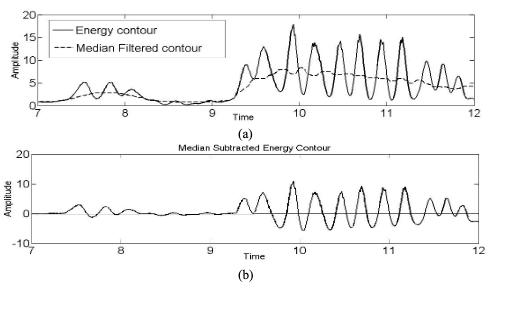}
\caption{Presence of tremolo in the energy contour of the \textit{taqsim} section of Turkish music as seen in (a) represented with a solid line and median filtered output represented in dashed line while (b) shows the periodic variation after subtraction of the energy contour from its median filtered output.} \label{fig6}
\end{figure}

\subsection{Microtonality based measure}
\begin{figure}
\includegraphics[width=\textwidth]{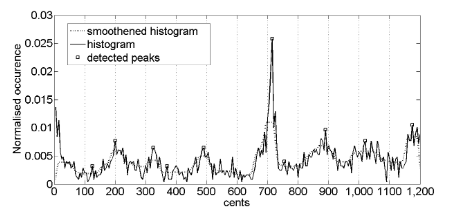}
\caption{Illustration of few steps in detecting peaks in a folded histogram for \textit{taqsim} section of \textit{makam} \textit{rast} by artist Hafiz Sesyilmaz (8 cent binning)} \label{fig7}
\end{figure}

Turkish music uses 53 Holdorein commas~\cite{bozkurt2011pitch} i.e. it has much more notes than in the case of Hindustani and Carnatic music. The presence of these micro-tones can serve to distinguish these tracks from Hindustani and Carnatic styles. Histogram of the pitch values can be used to find the location of the rendered notes from the expected note locations. The histogram of the notes in case of three styles shows that the Hindustani music notes are located maximum 10cents around the equitempered locations while the Carnatic music notes were seen to be lying around maximum 20cents of the equitempered note locations~\cite{koduri2012characterization}. In the case of Turkish music, the notes locations have seen to be off from the equitempered location by greater than 30cents~\cite{bozkurt2011pitch}. To come up with any feature based on the presence of micro-tones, we need to first find the peaks in the histogram of the audio. Unlike the approach of  Koduri et. al.~\cite{koduri2012characterization} or Bello et. al.~\cite{bello2005tutorial} for peak picking, where the histogram is smoother, in our case histogram is noisy due to the short duration of audio in the database. Thus in our case, the task of histogram peak picking becomes challenging. With this motivation, we look into modifying Bello et. al.’s approach for peak picking and characterize the behavior of three styles by coming up with four features. Instead of applying peak picking on the median subtracted histogram and then thresholding, we first we detect the peaks in the median filtered histogram. After detecting the peaks, we then go back to the original noisy histogram to search for the highest peak in the vicinity of 30cents around it as shown in  Figure \ref{fig7}. A similar approach was also used by Turnbull et. al for picking up boundaries for music segmentation in pop music~\cite{turnbull2007supervised}. 

We define the four features by making use of the equation below: 
\begin{equation}
    l'_{k}= f\left( l_{k}\right)= mod(l_{k},100), ~\hfill if ~\hfill l'_{k} <50, 
     ~\hfill else ~\hfill l'_{k}= 100-l'_{k}
\end{equation}                                                                                         
where $l_k$ is the location of the peak detected in the unfolded histogram 				
The $mod$ operation is carried out in order to remove the difference of two notes in terms of the distinction of the notes but retain the micro-tonal information. For eg., the difference between 110 cents and 290 cents will be the same as that from 790 cents.   
We convert the difference between the peaks greater than 50 cents to the range of 0 to 50 cents as a deviation for eg. of 80 cents ( Two peaks at 110 cents and 290 cents) is actually of 20 cents from the equitempered grid. 
The features described below are designed keeping in mind whether the tonic frequency of a particular track is available to us or not. Each of the feature requires the peak detection step as described above. Thus when there will be micro-tones present, the location of these peaks are expected to lie far from the equitempered grid. Even with the tonic information not present, we can exploit the fact that in some of the cases, there will be some notes having a considerable deviation from the equitempered grid.

\subsubsection{Feature design without the knowledge of the tonic of the piece}
This feature is designed for the cases where there is no explicit information about the tonic of the piece. We make use of the fact that only some of the notes are deviating from the equitempered grid. It is the maximum deviation of the peaks present in the histogram from each other which will capture the presence of any notes which have deviated from the equitempered grid.
A high value of the maximum inter peak deviation ($MIPD$) will indicate a strong presence of the micro-tones in an audio piece, where out of the $N$ peaks, we compute the maximum of the difference in locations $l_i$ and $l_j$.
\begin{equation}
    MIPD= max(f|l_i-l_j|) ,   i\neq j, i<j<N
\end{equation}

\subsubsection{Feature design with tonic of the piece known}
We propose three features if the tonic of the piece is known beforehand. They capture the deviation of the detected peaks from the pitch histogram from the equitempered grid obtained from the tonic information. We try to briefly describe the features below. The essence of these features is to capture the micro-tones rendered but in a different manner. 
Maximum peak deviation ($MPD$) calculates the maximum deviation of the peak location from the equitempered grid as 

\begin{equation}
    MPD = max(l'_{k})
\end{equation} 

We do not take the mean as the number of distinct peaks in a audio for a very small excerpt may vary according to the \textit{raga}/\textit{taqsim} being chosen. Instead of taking the maximum value we can assign a weight of the deviation by the height of the peaks. Thus weighted peak deviation ($WPD$) is calculated as to give more importance to the peak with higher note densities.
\begin{equation}
    WPD=\frac{\sum ^{N}_{k=1}f_{k}p_{k}}{N}
\end{equation}
									                   	 
We can find the micro-tonality of the rendered notes even without picking peaks in the histogram. This can be calculated by the equitempered note density feature ($ED$) which is the density of the notes lying in the vicinity of the equitempered location to the rest of the locations.
\begin{equation}
    ED=\frac{\sum_{k\in A }H_{k}}{\sum_{k}H_{k}}
\end{equation}

where $k \in [1,1200]$ and  $A$ = all $ k < dev$ from equitemptered scale.  We will select the final set of micro-tonality features by calculating the mutual information of features w.r.t each other to test whether there is a correlation present in the features or they are orthogonal to one another.

\section{FEATURE SELECTION BASED ON MUTUAL INFORMATION}
We carried out feature selection in two ways: 1. only within micro-tonality features (4.6) 2. considering all the features together(4.2-4.6). For both the cases feature selection was done considering information gain in WEKA toolbox giving a ranked list of features. From the 1st method, we selected the top two features among the four microtonality based features viz. maximum peak deviation and equitempered note location density. We can combine only the selected microtonality features thus obtained, along with other features to inspect the classification accuracy which we call feature subset $A$. 
Mutual information for a feature is calculated by:
\begin{equation}
   I(X_i,C)=H(C)-H(C|X_i) 
\end{equation}			                                                 
\begin{equation}
   H(C|X_i))=\sum_{c \in G}\sum_{x_i \in V_i}p(c,x_i) log(p(c|x_i))
\end{equation}

Here $X_i$ is the $i^{th}$ feature, $C$ is the class, $G$ is the set of classes, $V_i$ is the feature subset, $H$ is the entropy function, $p(c,x_i)$ and $p(c|x_i)$ is the joint and conditional distributions of the class and the $i^{th}$ feature respectively.

\begin{table*}[]
\resizebox{\textwidth}{!}{%
\begin{tabular}{|c|c|c|}
\hline
\textbf{Feature Subset A}                  & \textbf{Feature Subset B}             & \textbf{Feature Subset C}          \\ \hline
Equitempered note location density (ED) & Maximum Inter Peak Distance (MIPD) & MIPD                            \\ \hline
Maximum Peak Density (MPD)              & Equitempered location Density (ED) & MPD                             \\ \hline
Energy based feature                    & MPD                                & ED                              \\ \hline
Melodic Transitions                     & Energy based feature               & Energy based melodic transition \\ \hline
Distance of peak from tonic             & Melodic Transitions                & Distance of peak from tonic     \\ \hline
Gamak measure                           & Distance of peak from tonic        &   Melodic Transitions           \\ \hline
                                        & Gamak measure                      &                                 \\ \hline
\end{tabular}%
}
\caption{Features included in each subset}
\label{tab:FeatureSubsetNames}
\end{table*}

From the second method of feature selection, we inspect the accuracy by considering all features except the feature with least information gain (which we call feature subset $B$) and by considering all features except the last 2 features (which we call feature subset $C$). The list of all the features included in the subsets are presented in Table. 5. In feature subset $B$, we thus have excluded weighted peak deviation while in feature subset $C$ we have excluded Gamak measure in addition to weighted peak deviation. Note that in all the feature sets considered for information gain the microtonality based feature-weighted peak deviation is having least information gain. The accuracy is less in subset $B$ and $C$ as compared to $A$ as seen in Table \ref{tab:SubsetAccur}. The reason for the less accuracy in subset $B$ might be due to similar feature within the micro-tonality getting picked (maximum peak deviation and maximum interpeak difference are similar in the acoustic characteristic they are trying to capture). All the features are dealing with complementary properties in the audio except within the microtonality features. For example, energy-based features are orthogonal to other features with respect to the musicological concept that they are trying to capture. It is complementary to the micro-tonality based features which capture the consonance/dissonance felt by the listeners. Hence in the final model, we select feature subset $A$.

\begin{table*}[]
\centering
\begin{tabular}{|c|c|c|c|}
\hline
              & Feature Subset A & Feature Subset B & Feature Subset C \\ \hline
Accuracy (\%) & 85.5             & 84               & 83               \\ \hline
\end{tabular}
\caption{Accuracies after doing feature selection to obtain feature subsets A, B and C. The feature set A giving highest accuracy was chosen as the final model for classification.}
\label{tab:SubsetAccur}
\end{table*}

\section{AUTOMATIC CLASSIFICATION}
Classification was done using the feature set A using a discriminative and a generative classifier in a five-fold cross-validation mode. In addition to the melodic features, there exist timbral based differences due to the varied nature of instrumentation, language and the pronunciation in the three styles under consideration. These timbral differences can be modeled to a great extent by using MFCC coefficients which characterize the timbre of a sound. We used 13-MFCC coefficients averaged over the entire clip as the feature to capture the timbre with parameters chosen as given in CMU sphinx project using their implementation~\cite{CMUSphinx}. By making the stronger assumptions on the distribution of the data, the generative classifier will require lesser training data as compared to the discriminative classifier thus improving the classifier accuracy for the given dataset~\cite{ng2000cs229}. From the Table \ref{tab:ClassifierAccur}, it is observed that addition of our novel pitch based feature is improving the performance over the baseline timbre features. The classifier accuracy is highest for the quadratic classifier (using a full covariance matrix). Also, accuracy was compared for the various set of features namely (melodic features, timbre, and melodic+timbre based features) using the classifiers as seen in Table \ref{tab:ClassifierAccur}.

% \usepackage{graphicx}
\begin{table}[]
\centering
\resizebox{\textwidth}{!}{%
\begin{tabular}{|c|c|c|c|}
\hline
\textbf{Classifier} & \textbf{\begin{tabular}[c]{@{}c@{}}Accuracy \%\\ (Melody based Features)\end{tabular}} & \textbf{\begin{tabular}[c]{@{}c@{}}Accuracy \%\\ (Timbre based features)\end{tabular}} & \textbf{\begin{tabular}[c]{@{}c@{}}Accuracy \%\\ (Timbre+Melody based features)\end{tabular}} \\ \hline
Quadratic           & 85.5 (154/180)                                                                         & 85 (153/180)                                                                           & 92 (166/180)                                                                                  \\ \hline
kNN (k=7)           & 82 (148/180)                                                                           & 76.5 (138/180)                                                                         & 87.7 (158/180)                                                                                \\ \hline
\end{tabular}%
}
\caption{Accuracy using various sets of features for the discriminative (kNN) and generative (quadratic) classifier. Addition of melody based features to the baseline timbre based features shows improvement in the accuracy.}
\label{tab:ClassifierAccur}
\end{table}

The confusion matrix for the quadratic classifier can be seen in Table \ref{tab:ConfMatCV}. The confusion between the Carnatic and Turkish clips is more while the Hindustani class is well separated. As will be discussed, in the listening test section the confusion is seen more in Carnatic and Turkish styles in listeners as well. 

\begin{table*}[]
\centering
\begin{tabular}{c|c|c|c|}
\cline{2-4}
\multicolumn{1}{l|}{}               & \multicolumn{3}{l|}{\textbf{Observed}} \\ \hline
\multicolumn{1}{|l|}{\textbf{True}} & \textbf{C}  & \textbf{T}  & \textbf{H} \\ \hline
\multicolumn{1}{|c|}{\textbf{C}}    & 50          & 9           & 2          \\ \hline
\multicolumn{1}{|c|}{\textbf{T}}    & 8           & 51          & 1          \\ \hline
\multicolumn{1}{|c|}{\textbf{H}}    & 2           & 0           & 57         \\ \hline
\end{tabular}
\caption{Confusion matrix for classifier output using quadratic classifier in 5-fold cross-validation}
\label{tab:ConfMatCV}
\end{table*}

\section{CORRELATION OF LISTENING TEST AND CLASSIFIER LABELS}
\begin{figure}
\includegraphics[width=\textwidth]{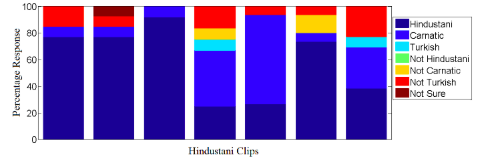}
\caption{Percentage responses across all the listeners of few Hindustani clips depicting the need for using two-best ratio over the percentage of the majority label as the confidence
} \label{fig8}
\end{figure}

We obtain the label of each track for the case of listening tests by taking the majority of the labels assigned by the listeners for the track across listener categories. The majority among the labels was always seen to be among H, C or T from the possible label options of H, C, T, NH, NC, NT, and NS. This eliminates the need for converting classifier labels into the 7 categories corresponding to the listening tests. As can be seen from the Figure \ref{fig1}, many of the clips have the second largest label marked by the listeners comparable to the first largest. Hence, just considering the majority label as the final label will not bring in the confidence of the listener markings. So, along with the label, we need its confidence value as well. One way to get the confidence is by just considering the percentage of the majority label. As seen from the Figure \ref{fig8}, this will be appropriate only in the case of the first three clips where a clear-cut majority is observed. Hence, we consider the ratio of the percentage of the highest marked label to the percentage of the second highest marked label as the confidence for the final label assigned. With this measure, comparing the 5th and the 6th clips, even though the percentage of the first majority label is equal, taking the two best ratios will clearly distinguish them. To bring the range of the confidence to 0-1 and avoid spurious peaks, we add 100 (i.e. highest percentage) to the denominator along with the percentage of the second highest label.  Label assigned to each track and its confidence is thus obtained for the listening tests as well as the classifier. These confidence values are then compared to perform statistical tests in order to draw corresponance between listening test confidence for each clip and that from automated classifier. 

We obtain a label as given by the classifier, by the output of a Quadratic generative classifier in a 5-fold cross validation mode, for each of the track from the database. This label is categorical i.e. it is an element of the discrete set 1, 0, -1 corresponding to Hindustani, Carnatic or Turkish classes respectively. As a measure of confidence of these labels, we also obtain the ratio of the posteriori probability of the most probable class to the second most probable class. Here again to bring the confidence values in the range of 0-1 and to avoid spurious high value, we add 1 (i.e. highest posterior probability value) in the denominator along with the second most probable class while taking the ratio. Thus for each of the tracks we now have a categorical label and a confidence value in the range of 0 to 1. In order to draw statistical correspondence between the outcome of the classifier and the perceptual listening test results, we perform two statistical tests, and compute the statistical significance for our hypothesis quantitatively. Two measures namely T-test and ANOVA were chosen for the same.

--T-test:
In our case the correlation coefficient came out to be 0.89 using Pearson correlation between the classifier labels and the listening test labels. The corresponding t-value is 26.19 for n=180 (samples). Comparing this t-value with the significance value derived from the t-table (with 5\% significance i.e here 1.96, (1.96<<26.19)) we conclude that the null hypothesis is getting rejected. So there is high correlation which is statistically significant for the current data size between the listening tests and classifier labels.

--Two way ANOVA with replication:
In our case we have used two way ANOVA with replication (56 replicates after outlier removal of 4 clips). First the contingency table is formed for the data for which the relationship needs to be found. This table is different from table 1, as it has labels along with their confidences of both listening tests and classifier. On this table a two way anova is carried out after removal of anomalies (Outlier removal) and normalization in case the data is originating from different methods. Matlab function was used to find the interaction between listening test and classifier. So the null hypothesis is that there is no interaction. This hypothesis was rejected implying that there is strong correlation between listening test and the classifier.

\section{CONCLUSION}

We have successfully proposed melodic features to distinguish Hindustani, Carnatic and Turkish Music. The addition of energy and microtonality based features have helped in distinguishing Turkish musical style from Indian Classical Music. Listening tests confirmed that the style distinction is evident in melody. The high correlation between the listening tests and the classifier output indicates the high dependence of the features on melodic cues. Moreover, the clips with high confidence of classification also correspond to the clips which all the listeners have marked to be of that particular style emphasizing the fact that the model chosen is able to distinguish the styles as per the cues used by the listeners. Addition of timbre based features over the melody based features shows considerable improvement to the baseline features. Misclassification between Carnatic and Turkish music is the highest and the addition of features modeling the non-steady regions might help to improve the accuracy further. 

\bibliographystyle{splncs04}
\bibliography{refer}

\end{document}